\documentclass[letter]{aa} 
\usepackage{graphicx}
\usepackage{txfonts}
\usepackage{natbib}
\bibpunct{(}{)}{;}{a}{}{,} 
\usepackage{color}
\def\mso{\,\mathrm{M}_\odot}
\def\rso{\,{\rm R}_\odot}

\newcommand{\gcm}{\,{\rm g}\,{\rm cm}^{-3}}

\def\gcms{\, {\rm g}\, {\rm cm}^{2}\, {\rm s}^{-1}}

\def\simle{\mathrel{\hbox{\rlap{\hbox{\lower4pt\hbox{$\sim$}}}\hbox{$<$}}}}
\def\simgr{\mathrel{\hbox{\rlap{\hbox{\lower4pt\hbox{$\sim$}}}\hbox{$>$}}}}

\begin{document}
\title{White dwarf spins from low mass stellar evolution models}
\author{M. P. L. Suijs\inst{1}, N. Langer\inst{1}, A.-J. Poelarends\inst{1},
           S.-C. Yoon\inst{2}, 
           A. Heger\inst{2,3}, F. Herwig\inst{4}}

  \institute{Astronomical Institute, Utrecht University, 
              P.O.Box 80000, 3508 TA, Utrecht, The Netherlands
             \and
              Department of Astronomy and Astrophysics, University
              of California, Santa Cruz, CA 95060, U.S.A.
              \and
              Theoretical Astrophysics Group, T-6, MS B227, Los
              Alamos, NM 87545, U.S.A.
              \and
              Keele Astrophysics Group, School of Physical and Geographical Sciences,
              Keele University, Staffordshire ST5 5BG, UK
	     }

\date{Received <date> / Accepted <date>}

\abstract   
              {The prediction of the spins of the compact remnants
               is a fundamental goal of the theory of stellar evolution.}
              {Here, we confront the predictions for white dwarf spins from 
               evolutionary models
               including rotation with 
               observational constraints.}
              {We perform stellar evolution calculations for stars in the mass range
            1\dots 3$\mso$, including the physics of rotation, from the zero age main
            sequence into the TP-AGB stage. We calculate
            two sets of model sequences, with and without inclusion of magnetic fields.
            From the final computed models of each sequence, 
            we deduce the angular momenta and
            rotational velocities of the emerging white dwarfs. }
            {While models including magnetic torques predict white dwarf rotational
            velocities between 2 and 10\,km\,s$^{-1}$, those from the non-magnetic
            sequences are found to be one to two orders of magnitude larger, well above
            empirical upper limits.}
            {We find the situation
            analogous to that in the neutron star progenitor mass range, and conclude
            that magnetic torques may be required in order to understand the
            slow rotation of compact stellar remnants in general.}

\authorrunning{Suijs et al.}
\titlerunning{White dwarf spins}
\maketitle


\section{Introduction}
During the last decade, rotation and rotationally
induced transport processes have been a major focus of 
the theory of massive star evolution (Maeder \& Meynet 2000,
Heger et al. 2000). Evolutionary models without 
magnetic field induced internal angular momentum transport
predict spins of newly born neutron stars which are 1...2 orders
of magnitude above those deduced from the youngest Galactic pulsars
(Ott et al. 2006). However, the magnetic torques proposed by Spruit (1998, 2002)
provide enough coupling between the rapidly rotating core and
the slowly rotating envelope in post-main sequence stars
to produce the right amount of core spin-down (Heger et al. 2005,
Petrovic et al. 2005). 

In {models of} low mass stars, 
the role of non-magnetic rotational transport processes for the determination
of the white dwarf spin is rather negligible, which leads to
rapidly rotating CO-cores
(Langer et al. 1999, Palacios et al. 2003, 2006; cf. below).
The huge shear above these rotating CO-core
during the thermally pulsing AGB phase inhibits the s-process
by swamping the $^{13}$C-pocket with the neutron poison $^{14}$N
(Herwig et al. 2003, Siess et al. 2004). 
Analogous to the situation in massive stars, it appears
that such rapidly spinning cores {in low mass star models} also contradict direct observations
of compact stellar remnants.
In a recent analysis of the Ca line profiles of a large sample of DA
white dwarfs, Berger et al. (2005) concluded that their rotational
velocities are generally below 10\,km\,s$^{-1}$, which is the smallest upper
limit derived from spectroscopy so far.  They conclude that the
predicted white dwarf spin from the rotating, non-magnetic evolutionary models
of Langer et al. (1999) can be ruled out. This conclusion is confirmed
by Kawaler (2004), who argues that the rotation rate of ZZ~Ceti pulsators
is even at least one order of magnitude slower than Berger et al.'s
upper limit. 
 
Internal magnetic torques have been
suggested as an agent to spin down the cores of white dwarf
progenitors during the giant stage (Spruit 1998). 
In this paper, we investigate whether magnetic torques, as 
computed in Spruit (2002), are able to alleviate the
problem of slowly spinning white dwarfs. Eggenberger et
al. (2005) already used the same physics to compute the
angular momentum evolution of a solar mass star, with the
result that the flat internal angular velocity profile
of our Sun could be recovered. 

Magnetic torques may not be the only mechanism to provide a
spin-down of the stellar core. E.g., Charbonnel \& Talon (2005)
invoked angular momentum transport though gravity waves 
(Zahn et al. 1997) to explain the slow rotation of the Solar core.  
However, here we concentrate on the magnetic torque mechanism,
as it gives promising results for massive stars and can be readily
applied to the low mass regime.

\section{Method}
We use a 1-D hydrodynamic stellar evolution code (Yoon et al., 2006, and
references therein). It includes diffusive mixing due to
convection, semiconvection (Langer et al. 1985) and thermohaline mixing
as in Wellstein et al. (2001).
This code includes the effect of the centrifugal
force on the stellar structure, and time-dependent chemical mixing and
transport of angular momentum due to rotationally induced
instabilities (Heger et al. 2000).  We also include
chemical mixing and transport of angular momentum due to magnetic
fields (Spruit 2002) as in Heger et al. (2005) and Petrovic et al. (2005).
We assume rigid rotation in convection zones.

We calculate the evolution of solar metallicity stars starting 
rigidly rotating at the zero-age main
sequence, for initial masses of 1.0$\mso$, 1.5$\mso$, 2.0$\mso$ and
3.0$\mso$, with and without the effects of magnetic torques. The initial
equatorial velocities of these models were chosen to be 2, 45, 140
and 250\,km\,s$^{-1}$ (Tassoul, 2000). For the solar mass models, we assume that the 
angular momentum loss due to magnetic braking connected with the solar-type
wind is already over
once we start our calculations (cf. Eggenberger et al. 2005).
All but our 3$\mso$ models evolve through the core helium flash.
Throughout the evolution of all models, the mass loss
rate of Reimers (1975) was used. 
The choice of the AGB mass loss rates does
not affect our results significantly.  The calculations were stopped
after at least five and at most 28 thermal pulses, {beyond which the
CO-core angular momentum is not expected to change significantly (cf., Sect.~3)}.

We compute the surface rotational velocities of the white dwarfs emerging from
our model sequences as follows. First, we fix the final white dwarf mass
according to the initial-final mass relation from Weidemann (2000). These
white dwarf masses (cf. Table 1) are generally slightly larger than the final CO-core
masses of our models. However, as we shall see below, the total angular momentum
of the CO-core does not change any more during its growth on the TP-AGB,
and therefore, even though we correct our final CO-core
masses in the mentioned way, our final angular momenta do not need to
be adjusted. 

The surface rotational velocities are then calculated assuming rigid rotation
inside the white dwarf. This is well justified due to the short
timescale of angular momentum redistribution ($\sim 10^8\,$yr) inside even the non-magnetic
models (Yoon \& Langer 2004). We then use the gyration radii, $H$,
for polytropes derived by Motz (1952).  For a white dwarf with a polytropic
index $n=1.5$, mass $M_\mathrm{WD}$, radius $R_\mathrm{WD}$ ({we employ the
white dwarf mass-radius relation of} Hamada \& Salpeter, 1961), and
angular momentum $J_\mathrm{tot}$, it is (Table 1 in
Motz, 1952):
\begin{equation}
\left(\frac{H}{R_\mathrm{WD}}\right)^{2} = 0.20502\;.
\end{equation}
With $\omega=J/I$ and
$I_\mathrm{WD}=H^2 M_\mathrm{WD}$
we obtain the surface velocity
\begin{equation}
v_\mathrm{rot,WD} = \omega_\mathrm{WD} R_\mathrm{WD} = \frac{J_\mathrm{tot}}{I_\mathrm{WD}} R_\mathrm{WD} = \frac{J_\mathrm{tot}}{0.20502 R_\mathrm{WD} M_\mathrm{WD}},
\end{equation}
\label{eq-v}
where $\omega_\mathrm{WD}$ is the angular velocity of the white dwarf.
We have tested this procedure by computing rotating cold white dwarf models
with our stellar evolution code and find it accurate to within 10\% 
for the slowly rotating cases considered here. 


\section{Results}

\begin{figure}[]
 \includegraphics[width=0.72\hsize,angle=-90]{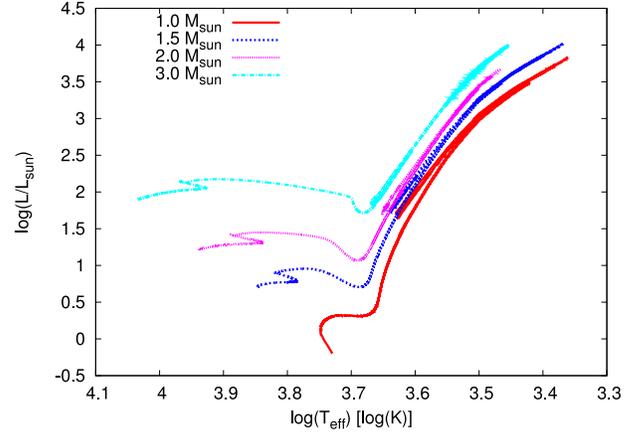}
 \caption{Evolutionary tracks of the four sequences including magnetic torques 
in the Hertzsprung-Russell diagram (cf. Table~1).
 	}
 \label {fig-4HR}
\end{figure}

\begin{figure}[]
 \includegraphics[width=0.72\hsize,angle=-90]{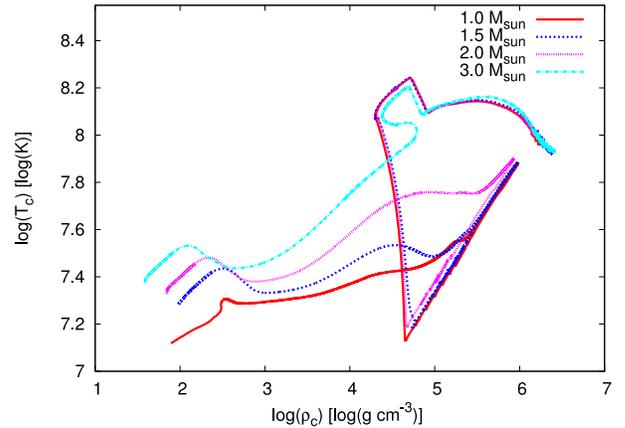}
 \caption{Evolution of the four sequences including magnetic torques in the
central density ($\rho_\mathrm{c}$) versus central
   temperature (T$_\mathrm{c}$) plane}
 \label {fig-4rho}
\end{figure}

The evolutionary tracks of our four magnetic models in the HR-diagram are shown in Fig.~\ref{fig-4HR},
while Fig.~\ref{fig-4rho} displays their evolution in the $T_c-\rho_c$-plane.
The tracks of our non-magnetic sequences would be indistinguishable on these plots.
Fig.~\ref{fig-4rho} shows that during the late AGB evolution, the CO-cores of our models
are degenerate and cooling. Their central densities are close to that of cold white dwarfs 
in the the considered mass range, i.e. $\rho_{\rm c}\simeq 3...6\, 10^6\gcm$ 
for $M_{\rm WD}=0.55\mso ... 0.7\mso$. 

In order to visualize the angular momentum distribution in our models,
we show diagrams using the integrated angular
momentum $J(M_r)=\int_{\,0}^{M_r}j(M_r)\mathrm{d} M_r$ divided by
$M_{r}^{5/3}$, as a function of the mass coordinate $M_r$.  
While a homogeneous, rigidly rotating body results in a horizontal line
in such a diagram,
the $J(M_r)$-profiles from different evolutionary stages trace the 
flow of angular momentum through the mass shells, since $J(M_r)$ and
$J(M_r) M_r^{-5/3}$ remain constant in a given mass shell if no
angular momentum is transported through this shell.

\begin{figure}[]
 \includegraphics[width=0.72\hsize,angle=90]{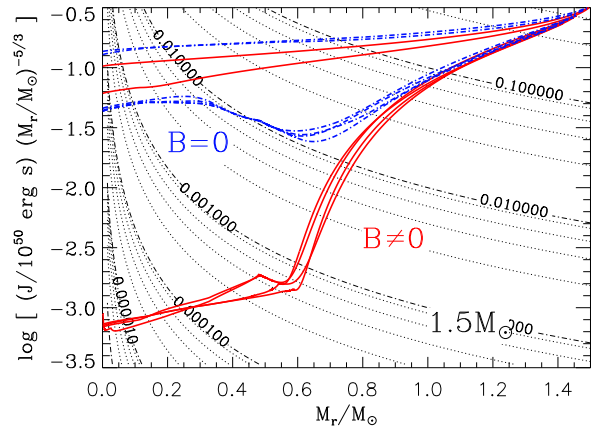}
 \caption{Integrated angular momentum $J(M_r)
   =\int_{0}^{M_r}j(m)\mathrm{d} m$ divided by $M_{r}^{5/3}$ as a
   function of the mass coordinate, at different times for the magnetic 
   (full drawn red lines) and
   non-magnetic (dashed blue lines) 1.5$\mso$ sequences. The contours display levels of
   constant $J$, labeled with log($J/10^{50}$erg~s). The values for
   the dotted contours are 3, 5, 7, and 9 times the value of the
   labeled contour below them. 
   From high to low, the six lines per sequence represent the
   following evolutionary stages: start main sequence (X$_c=0.7$),
   main sequence (X$_c\sim0.25$), start core He burning (Y$_c\sim0.95$),
   He burning (Y$_c\sim0.4$), start TP phase and during TP phase ($>$5
   pulses). }
 \label {fig-J1.5}
\end{figure}

Fig.~\ref{fig-J1.5} displays $J$-profiles from six evolutionary
stages for our 1.5$\mso$ model, both with and without the effects of
magnetic torques. For the non-magnetic model, the deepest extension
of the convective envelope is down to about $0.2\mso$. Therefore,
the helium core is built-up starting with the initial angular momentum
of the inner $~0.2\mso$, i.e. about $10^{48}\,$cm$^2$\,s$^{-1}$.
During the growth of the helium core on the RGB, material from the
former bottom of the convective envelope is incorporated,
which is very slowly rotating and which, in its inner parts,
has a specific angular momentum several orders of magnitude below that of
the material in the helium core. Thus, effectively, the helium core
grows in mass without gaining angular momentum. This feature is continued
for the CO-core during the AGB-evolution. As a consequence, the $J$-profiles
in Fig.~\ref{fig-J1.5} follow the line of $J=10^{48}\,$cm$^2$\,s$^{-1}$
in the mass range $0.2\mso < M_{\rm r} < 0.6\mso$ during the post main
sequence evolution. Other processes than convection are clearly negligible
in this case. The specific angular momentum in the inner $~0.6\mso$ core
of this model has been lowered by about a factor~6, from the ZAMS to the
white dwarf stage. For the other non-magnetic models, this factor is 
rather similar.


\begin{table}

	\caption{
	Initial mass $M_\mathrm{i}$,  
        initial equatorial velocity $\varv_{\mathrm{rot,i}}$,
        adopted white dwarf mass $M_{\mathrm{WD}}$ and radius $R_{\mathrm{WD}}$ 
        final core angular momentum $J_\mathrm{c,f}$, 
        final mass averaged specific core angular momentum $j_{\mathrm{c,f}}$
        and expected equatorial rotation velocity ($\varv_{\mathrm{rot,WD}}$) from Eq.~2,
        for our magnetic (upper 4 lines) and non-magnetic models.
		}
	\label{tab-values}
	
	\begin{tabular*}{\linewidth}{@{\extracolsep{\fill}}ccccccc}
	\hline
	\noalign{\smallskip}
	$M_{\mathrm{i}}$  & $\varv_{\mathrm{rot,i}}$  & $M_{\mathrm{WD}}$ & 
        $R_{\mathrm{WD}}$ & $J_{\mathrm{c,f}}/10^{46}$ & 
        $j_{\mathrm{c,f}}/10^{13}$ & $\varv_{\mathrm{rot,WD}}$  \\
        $\mso$ &  $\, {\rm km}\, {\rm s}^{-1}$ &  $\mso$ & $10^{-2}\,\rso$ & $\gcms$ & $\,{\mathrm{cm}^2\, {\rm s}^{-1}}$ & 
        $\, {\rm km}\, {\rm s}^{-1}$ \\

	\noalign{\smallskip}
	\hline
	\noalign{\smallskip}
                                                                                                

        1.0 & 2   & 0.550 & 1.25 & 5.00 & 4.54 & 2.6 \\
        1.5 & 45  & 0.575 & 1.20 & 5.83 & 5.07 & 3.0 \\
        2.0 & 140 & 0.600 & 1.19 & 8.05 & 6.71 & 4.5 \\
        3.0 & 250 & 0.680 & 1.08 & 15.4 & 11.3 & 9.1 \\
        \noalign{\smallskip}
        1.0 & 2   & 0.550 & 1.25 & 8.07 & 7.33 & 4.2 \\
        1.5 & 45  & 0.575 & 1.20 & 110  & 95.6 & 56  \\
        2.0 & 140 & 0.600 & 1.19 & 329  & 274  & 180 \\
        3.0 & 250 & 0.680 & 1.08 & 458  & 336  & 220 \\

        \noalign{\smallskip}
        \hline
	\end{tabular*}

\end{table}

\begin{figure}[]
  \includegraphics[width=0.72\hsize,angle=90]{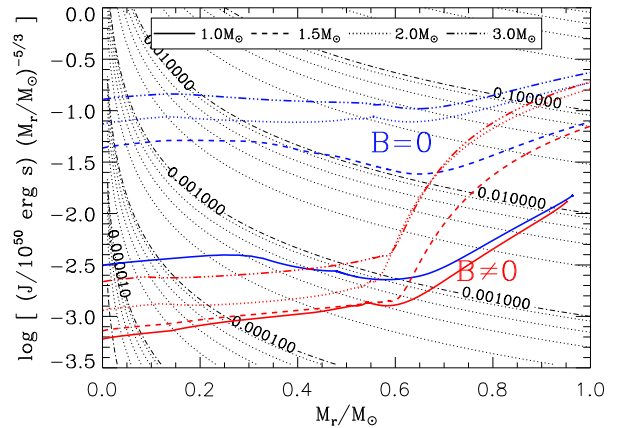}
  \caption{Integrated angular momentum $J(M_r)
    =\int_{0}^{M_r}j(m)\mathrm{d} m$ divided by $M_{r}^{5/3}$, as a
    function of the mass coordinate for the final stellar models of our sequences
    (cf. Table~1). The background contours display levels of constant $J$, as in Fig.~3.
     }
 \label {fig-finalJ}
\end{figure}

For the magnetic $1.5\mso$ sequence, Fig.~\ref{fig-J1.5} shows
a drop of core angular momentum already during core hydrogen burning.
This is a consequence of the magnetic fields enforcing close-to-rigid
rotation. However, the main drain of core angular momentum occurs
between core hydrogen exhaustion and core helium ignition,
i.e. during the RGB phase. The total loss of core angular momentum here
is about a factor of 20 larger than in the non-magnetic case, i.e.
a factor of 120 in total. Fig.~\ref{fig-finalJ}, which compares the
final angular momentum distributions of our magnetic and non-magnetic
models, shows that the situation is similar for all the studied 
cases, except for the solar mass models, where the initial angular momentum
was already rather low.

\section{Discussion}
\label{discussion}

\begin{figure}[]
  \begin{center}
      \includegraphics[width=0.75\hsize,angle=270]{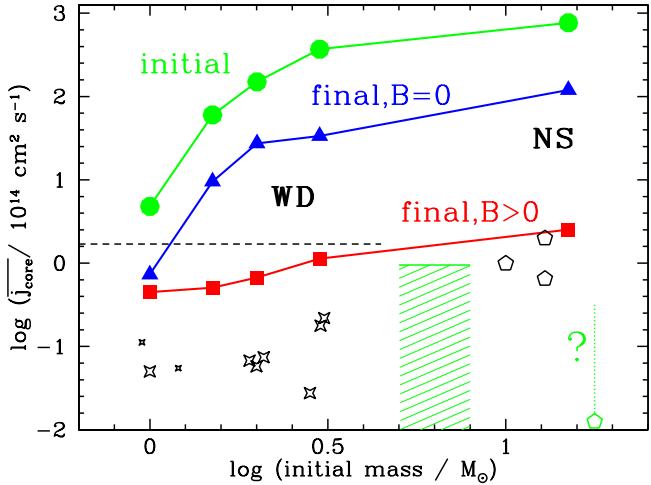}
       \caption{
         Average core specific angular momentum
         of our initial and final models (cf. Table~1 and Fig.~4), 
        versus initial stellar mass (full drawn lines).
        Points for 15$\mso$ stars computed with the same physics 
        from Heger et al. (2005) have been 
        added. The upper line corresponds to the initial models.
        Filled triangles corresponds to the final
        models of the non-magnetic sequences, and filled squares
        to the final models of the magnetic sequences. 
        The dashed horizontal line indicates the spectroscopic upper limit 
        on the white dwarf spins obtained by Berger et al. (2005).
        Star symbols represent astroseismic measurements from ZZ Ceti stars
        (Bradley 1998, 2001; Dolez 2006; Handler 2001, Handler et al. 2002, Kepler et al. 1995, 
        Kleinmann et al. 1998, Winget et al. 1994),
        where smaller symbols correspond to less certain measurements.   
        The green hatched area is populated by magnetic white dwarfs
        (Ferrario \& Wickramasinghe 2005; Brinkworth et al. 2007).
        The three black open pentagons correspond to the youngest Galactic 
        neutron stars (Heger et al. 2005), while the green pentagon is
        thought to roughly correspond to magnetars (Camilo et al. 2007), where  
        the vertical dotted green line indicates the possibility that
        magnetars are born with higher core angular momentum.
        }
	\label{fig-specj}
  \end{center}
\end{figure}

Fig.~\ref{fig-specj} summarizes our results by showing the initial and the final
average specific core angular momentum of our model. It also
contains the corresponding result form Heger et al. (2005) for a 15$\mso$ star.
Observational constrains from white dwarf spectroscopy, pulsation analyses
of ZZ~Ceti stars, magnetic white dwarfs, and young neutron stars are also plotted
on this figure. 

In the initial mass range considered here, as well as in the massive star regime,
it is clear that the non-magnetic models fail by a large margin to comply
with the observations. Only for solar mass stars, where magnetic braking due to
the solar-type wind leads to a low initial angular momentum, could the non-magnetic
model be marginally consistent with the empirical data. However, our assumption was
that the slow surface rotation of these stars early during hydrogen burning 
has slowed down the rotation of the whole stellar interior. Eggenberger et al. (2005)
showed that exactly this might have been performed by the B-fields produced due
to the Spruit (2002) dynamo. 

The magnetic low mass models all fall below the spectroscopic limit on white dwarf
rotation of Berger et al. (2005). Towards more massive stars, the agreement with
observations seems good. The neutron star spins are well recovered, and the range
of spins of the magnetic white dwarfs (which are presumed to have
more massive progenitors than typical white dwarfs; Ferrario \& Wickramasinghe 2005)
reaches up to the line connecting 3$\mso$ and 15$\mso$. The partly
much slower rotation of the magnetic white dwarfs
could be explained by additional magnetic braking due to
stellar winds in the white dwarf progenitors.  
However, our magnetic models still predict white dwarf rotation rates
in the 1$\mso$ to 3$\mso$ range which are about one order of magnitude
larger than what is found in pulsating white dwarfs (ZZ~Ceti stars). 
This implies that the white dwarfs should lose additional angular momentum,
either after they are born, or still inside their progenitor stars.

In the first case, one may speculate whether the situation in analogous to the
neutron star case: for pulsars it is well known that they
spin down with time. As the ZZ~Ceti stars have a cooling age of 
about $10^9\,$yr, they might have lost angular momentum on their 
cooling track. However, estimating the spin-down time through a magnetic
wind by $\tau_1 = \dot M^{-{1\over 2}} (M/BR) (2GM/R)^{1\over 4}$ 
(Justham et al. 2006) would allow for a significant effect at most
in strongly magnetic white dwarfs, and spin-down through electromagnetic radiation,
with $\tau_2= M c^3 / (B^2 \Omega^2 R^4)$ is always insignificant.
In the above expressions, $M, R, \Omega$ and $B$ are white dwarf mass, radius,
surface magnetic field strength and spin frequency.

In the second case, perhaps an additional internal angular momentum
transport mechanism in white dwarf progenitors is required. 
While transport through gravity waves
has been suggested as such (Zahn et al. 1997), we point out that,
for stars above $\sim 1.3\mso$, the core spin-down is likely to occur 
during the post-main sequence evolution, since on the main sequence
these stars are still rapid rotators. It is unclear at present whether
gravity waves can transport angular momentum through the shell source
in the giant stage. On the other hand, the magnetic angular momentum transport 
has been shown by Zahn et al. (2007) to be more complex than in the picture
of Spruit (2002). 
Also, Yoon et al. (2008) find strong poloidal fields may be generated
by convective cores, and suggest that their influence on angular momentum
transport may be comparable to that of the fields suggested by Spruit.


\section{Conclusions}

While nature may be more complex, it is worthwhile to attempt to understand
the spins of compact stellar remnants with a single theoretical approach
for all initial masses. 
We showed above that angular momentum transport through
rotationally induced magnetic fields according to Spruit (2002) 
provides a major improvement of the predicted spins of white dwarfs
and neutron stars. Magnetic angular momentum transport appears also
the most promissing candidate to bridge the still existing gap
between observed and predicted white dwarf spins at low mass (see Fig.~5).

If internal magnetic torques in their progenitors are indeed 
resopnsible for the slow rotation of compact stars in the Milky Way, then 
there may be little room for an important role of angular momentum in their
formation process, at least in single stars. This may have implications 
for understanding the progenitors and the formation meachanism of
magnetars (Sawai et al. 2008) and long gamma-ray bursts (Woosley 
and Bloom 2006).

%

\begin{acknowledgements}
AH performed this work under the auspices of the National Nuclear
Security Administration of the U.S. Department of Energy at Los Alamos
National Laboratory under Contract No. DE-AC52-06NA25396,  
{and was supported by the DOE Program for
Scientific Discovery through Advanced Computing (SciDAC;
DE-FC02-01ER41176)}.
\end{acknowledgements}


\begin{thebibliography}{}


\bibitem[]{} Berger, L., Koester, D., Napiwotzki, R., Reid, I.~N., Zuckerman, B., 2005, A\&A, 444, 565
\bibitem[]{} Bradley, P. A., 1998, ApJS 116, 307
\bibitem[]{} Bradley, P. A., 2001, ApJ 552, 326
\bibitem[]{} Brinkworth, C.S., Burleigh, M.R., Marsh, T.R., 2007, in: 
   15th European Workshop on White Dwarfs, ASP Conference Series, Vol. 372, San Francisco, 
   Ralf Napiwotzki and Matthew R. Burleigh, eds., p.183
\bibitem[]{} Camilo, F., Ransom, S.M., Halpern, J.P., Reynolds, J., 2007, ApJ, 666, L93
\bibitem[]{} Charbonnel, C., Talon, S., 2005, Science, 309, 2189
\bibitem[]{} Dolez N., Vauclair, G., Kleinman, S. J., et al., 2006, A\&A 446, 237 
\bibitem[]{} Eggenberger, P., Meynet, G., Maeder, A., 2005, A\&A, 440, L9
\bibitem[]{} Ferrario, L., Wickramasinghe, D. T., 2005, MNRAS, 356, 615
\bibitem[]{} Hamada, T., Salpeter, E.E., 1961, ApJ, 134, 683
\bibitem[]{} Handler, G., 2001, MNRAS, 323, L43
\bibitem[]{} Handler, G., Romero-Colmenero, E., Montgomery, M. H., 2002, MNRAS, 335, 399 
\bibitem[]{} Heger, A., Langer, N., \& Woosley, S.E., 2000, A\&A, 2000, ApJ, 528, 368
\bibitem[]{} Heger, A., Woosley, S.E., \& Spruit, H.C., 2005, ApJ, 626, 350
\bibitem[]{} Justham, S.; Rappaport, S.; Podsiadlowski, P., 2006, MNRAS, 366, 1415
\bibitem[]{} Kawaler S.D., 2004, in: IAU-Symp. No.~215 on ``Stellar Rotation'', A.~Maeder and
   P.~Eenens, eds, ASP, p. 561
\bibitem[]{} Herwig, F., Langer, N., Lugaro, M., 2003, ApJ, 593, 1056
\bibitem[]{} Kepler, S.O., Giovannini, O., Wood, M.A., et al., 1995, ApJ 447, 874
\bibitem[]{} Kleinman, S. J.; Nather, R. E.; Winget, D. E., et al., 1998, ApJ 495, 424
\bibitem[]{} Langer, N., El Eid M.F., Fricke K.J., 1985, A\&A, 145, 179
\bibitem[]{} Langer, N., Heger, A., Wellstein, S., Woosley, S.E., 1999, A\&A, 346, L37
\bibitem[]{} Maeder, A., Meynet, G., 2000, ARAA, 38, 143
\bibitem[]{} Maeder, A., Meynet, G., 2005, A\&A, 440, 1041
\bibitem[]{} Motz, L., 1952, ApJ, 115, 562
\bibitem[]{} Ott, C. D.; Burrows, A.; Thompson, T. A.; Livne, E.; Walder, R., 2006, ApJS, 164, 130
\bibitem[]{} Palacios, A., Talon, S., Charbonnel, C., Forestini, M., 2003, A\&A, 399, 603
\bibitem[]{} Palacios, A., Charbonnel, C., Talon, S., Siess, L., 2006, A\&A, 453, 261
\bibitem[]{} Petrovic, J., Langer, N., Yoon, S.-C., Heger, A., 2005, A\&A, 435, 247
\bibitem[]{} Reimers, D., 1975, Mem. Soc. Liege, 8, 369
\bibitem[]{} Sawai, H.; Kotake, K.; Yamada, S., 2008, ApJ, 672, 465
\bibitem[]{} Siess, L.; Goriely, S.; Langer, N., 2004, A\&A, 415, 1089
\bibitem[]{} Spruit, H.C., 1998, A\&A, 333, 603
\bibitem[]{} Spruit, H.C., 2002, A\&A, 381, 923 
\bibitem[]{} Tassoul, J.L., 2000, {\it Stellar Rotation}, Cambridge University Press
\bibitem[]{} Weidemann, V., 2000, A\&A, 363, 647
\bibitem[]{} Wellstein S., Langer N., Braun H., 2001, A\&A, 369, 939
\bibitem[]{} Winget, D.E., Nather, R.E., Clemens, J. C., 1994, ApJ 430, 839
\bibitem[]{} Woosley, S.E., Bloom, J.S., 2006, ARAA, 44, 507
\bibitem[]{} Yoon, S.-C., Langer, N., 2004, A\&A, 419, 623
\bibitem[]{} Yoon, S.-C., Langer, N., Norman, C., 2006, A\&A, 460, 199
\bibitem[]{} Yoon, S.-C., et al., Proc. IAU-Symp. 250, in press
\bibitem[]{} Zahn, J.-P.; Talon, S.; Matias, J., 1997, A\&A, 322, 320
\bibitem[]{} Zahn, J.-P.; Brun, A. S.; Mathis, S., 2007, A\&A, 474, 145
\end{thebibliography}
\end{document}